\documentclass[11pt]{scrartcl}

\addtokomafont{caption}{\footnotesize}
\addtokomafont{captionlabel}{\bfseries\sffamily}
\setcapindent{0em}

\usepackage{manuscript_packages}
\usepackage[title]{appendix}

\usepackage[pagewise]{lineno}

\usepackage{pdfpages}

\usepackage{todonotes}
\usepackage{ulem}
\renewcommand{\emph}[1]{\textit{#1}}

\usepackage{subcaption}

\usepackage{xcolor}

\usepackage{hyperref}
\hypersetup{
	hidelinks = true,
	colorlinks = false,
	linkcolor = {red!80!black},
	anchorcolor = {black},
	citecolor = {green!80!black},
	filecolor = {cyan!80!black},
	menucolor = {red!80!black},
	runcolor = {cyan!80!black},
	urlcolor = {magenta!80!black},
}

\newabbreviation{discrete dislocation dynamics}{DDD}{discrete dislocation dynamics}
\newabbreviation{face-centered cubic}{FCC}{face-centered cubic}
\newabbreviation{coefficient of variation}{CV}{coefficient of variation}
\newabbreviation{mean absolute deviation}{MAD}{mean absolute deviation}
\newabbreviation{discrete-to-continuous}{D2C}{discrete-to-continuous}
\newabbreviation{continuum dislocation dynamics}{CDD}{continuum dislocation dynamics}
\newabbreviation{kernel density estimation}{KDE}{kernel density estimation}

\title{Statistical analysis of Discrete Dislocation Dynamics simulations: initial structures, cross-slip and microstructure evolution}

\begin{document}
	

\author[1]{Aytekin Demirci}
\author[2]{Dominik Steinberger}
\author[3]{Markus Stricker}
\author[2,4]{Nina Merkert}
\author[5]{Daniel Weygand}
\author[1, 2, 6, *]{Stefan Sandfeld}
\affil[1]{Institute for Advanced Simulations -- Materials Data Science and Informatics (IAS-9), Forschungszentrum Juelich, Juelich, Germany}
\affil[2]{Institute of Mechanics and Fluid Dynamics, Freiberg University of Mining and Technology, Freiberg, Germany}
\affil[3]{Interdisciplinary Centre for Advanced Materials Simulation, Ruhr-Universität Bochum, Bochum, Germany}
\affil[4]{Institute of Applied Mechanics, Clausthal University of Technology, Clausthal-Zellerfeld, Germany}
\affil[5]{Institute for Applied Materials, Karlsruhe Institute of Technology, Karlsruhe, Germany}
\affil[6]{RWTH Aachen University, Faculty of Georesources and Materials Engineering, Chair of Materials Data Science and Materials Informatics, Aachen, Germany}
\affil[*]{Corresponding author: s.sandfeld@fz-juelich.de}

\maketitle





%
%

\section{Abstract}
Over the past decades, discrete dislocation dynamics simulations have been shown to reliably predict the evolution of dislocation microstructures for micrometer-sized metallic samples. Such simulations provide insight into the governing deformation mechanisms and the interplay between different physical phenomena such as dislocation reactions or cross-slip. This work is focused on a detailed analysis of the influence of the cross-slip on the evolution of dislocation systems. A tailored data mining strategy using the ``\ab*{discrete-to-continuous} framework'' allows to quantify differences and to quantitatively compare dislocation structures. We analyze the quantitative effects of the cross-slip on the microstructure in the course of a tensile test and a subsequent relaxation to present the role of cross-slip in the microstructure evolution. The precision of the extracted quantitative information using D2C strongly depends on the resolution of the domain averaging. We also analyze how the resolution of the averaging influences the distribution of total dislocation density and curvature fields of the specimen. Our analyzes are important approaches for interpreting the resulting structures calculated by dislocation dynamics simulations.

\section{Introduction}

Dislocations are one-dimensional defects found in crystalline materials. They
are the boundary of an area over which relative slip occurred
on defined slip planes~\autocite{Anderson_2017_a}. The crystal lattice
in the vicinity of the dislocation core is distorted, which results in long-range
stresses in the material. Dislocation glide, i.e., the expansion or contraction of the slipped areas is their response to the local stresses which is the sum of external loading and the stress field of other dislocation and results in the plastic deformation of the crystal.
But dislocations do not only interact via their respective stress field.
Their behavior is more complex and includes several types of topological changes: dislocations can form junctions to lower their elastic energy (\textit{Frank's rule}~\autocite{Hull2011}) or can change their glide plane through a process called cross-slip.
This leads to complex dislocation networks during straining of a specimen because, depending on individual dislocation's properties like slip plane and Burgers vector, these drivers of the topological changes can be mobile or immobile.
In this division of possible dislocation interactions, the process of cross-slip is of the mobile sort and thereby provides an additional path for the material to relax external loading.
While it has been accepted that cross-slip significantly impacts the formation of the dislocation microstructure, the exact impact of cross-slip onto the actual dislocation microstructure is yet to be quantified, especially as we are not able to arbitrarily \enquote{turn it on or off} in experiments.

\Ab*{discrete dislocation dynamics}~\autocites{Weygand_2002_a,Po_2014_c,LeSar_2020_a} allows to study the effect of particular mechanisms on the plastic deformation behavior due to dislocation propagation, e.g., how different junction types contribute to the strain hardening behavior of \ab*{face-centered cubic} crystals~\autocites{Weygand_2014,Stricker_2015_a}{Sills_2018_a}.
Using \ab*{discrete dislocation dynamics}, the impact of cross-slip on the stress-strain curve and the total dislocation density has been studied~\autocites{Motz_2009_a}{Zhou_2010_a}{Hussein_2015_a}.
A common observation is that cross-slip results in lower stresses and higher total dislocation densities for the same macroscopic strain when compared to the same numerical experiments without cross-slip.
Cross-slip was also identified as one of two processes providing \textit{new} dislocations~\autocite{Stricker_2018_a}.
Dislocation \enquote{network} characteristics were further studied in terms of the density of dislocation junctions~\autocite{Hussein_2015_a}.
Based on \ab*{discrete dislocation dynamics} simulations,~\textcite{Xia_2016_a} extracted cross-slip rates and used them to enhance a continuum dislocation dynamics model.




In the aforementioned work, the influence of cross-slip on the microstructure
was only considered in the sense of global densities of either the dislocations
themselves or of junctions formed by the dislocations.
This work puts the focus on \emph{where} dislocation microstructures are affected by cross-slip during uniaxial tension loading and unloading of a cubiod specimen on the micron scale.
Recent studies of dislocation microstructures from \ab*{discrete dislocation dynamics} simulations indicate that, given a moderately high dislocation density, dislocation motion and therefore plasticity is a relatively local phenomenon~\autocites{Stricker_2018_a}{Sudmanns2019}.
Therefore the study of \emph{where} dislocations interact is needed to contribute to the existing purely averaged approaches. 

Another open question related to dislocation microstructures apart from the
evolution is the initial state of a simulation.
Real specimen have an existing, physically consistent microstructure but simulations need to be initialized with a dislocation microstructure.
Ideally, one would use initial microstructures which are a statistically equivalent to real ones or at least equivalent to microstructures.
On a side note, this also raises the question of what a \emph{statistically equivalent} microstructure is and is connected to our second question.
Although there are several novel techniques and approaches for observing 3D structure of dislocations via experiments ~\autocites{LIU20141}{OVEISI2018116}{leon2020three}{steinberger2023data}{zhang2022data}, the precision of the state-of-art techniques is limited, therefore quantitative microstructure information can be accessed is insufficient to describe a complete dislocation network of a microstructure. Furthermore, initializing the microstructures is also a concern for \ab*{continuum dislocation dynamics} simulations where the initial variables other than the dislocation density are less amenable to “guesswork”. Both \ab*{discrete dislocation dynamics} and  \ab*{continuum dislocation dynamics} simulations require local statistics about how density or curvature changes during the deformation path and exploring the role of the cross-slip on those observables for a physically reasonable initialization of a microstructure. This is another question we explore here via \ab*{continuum dislocation dynamics} field variables obtained from coarse graining of \ab*{discrete dislocation dynamics} simulations microstructures.

Our study aims to perform a statistical characterization of discrete dislocation microstructures via \ab*{continuum dislocation dynamics} field variables. The reason of using continuous field variables is that \ab*{continuum dislocation dynamics} variables allow to base plasticity on dislocation-related measures: spatially resolved densities and curvatures, while we can access only simple statistics such as average dislocation densities via \ab*{discrete dislocation dynamics} simulations. In the following, we first introduce how we used large-scale \ab*{discrete dislocation dynamics} simulations to generate dislocation microstructures in uniaxial tensile tests up to a strain of \SI{0.6}{\percent} and subsequent unloading.
We do this twice for each initial dislocation microstructure; once with cross-slip and once without it.
Subsequently, we summarize the so-called \ab*{discrete-to-continuous} method~\autocites{Sandfeld_2015_c}{Steinberger_2016_b} and use it to convert the discrete dislocation to continuous field data. In this study, the fields which we make use of are total dislocation density and curvature fields. In general, the total dislocation density is important for direct quantitative comparison of coarse-grained microstructures, while the latter carries topological properties of dislocation structures~\autocite{weger2021analysing}. We then make use of D2C to compare how dislocation microstructures that form with and without cross-slip differ over the course of the loading and unloading.
The statistics of the unloaded dislocation microstructures are then analyzed and the resulting implications for initializing simulations are discussed.
Finally, we summarize and discuss the results.

\section{Methods}

\subsection{Discrete dislocation dynamics}
Throughout this work, we use the \ab*{discrete dislocation dynamics} code
described in \textcites{Weygand_2002_a, Weygand_2001_a} to generate dislocation
microstructures similar to the ones of \textcites{Motz_2009_a}{Stricker_2018_a}.
Material parameters for \ab*{face-centered cubic} aluminum are used:
The lattice constant is \SI{0.4045}{\nano\meter}, shear modulus of \SI{27}{\giga\pascal}, Poisson's ratio is \SI{0.347}{}.
The cuboid-shaped simulation box with free surfaces has a volume of 5x5x\SI{5}{\micro\meter}$^3$.  Its axes align with the crystallographic axes of the material, i.e., the $x$-axis is parallel to \hkl[100], the $y$-axis to \hkl[010], and the $z$-axis to \hkl[001].

\begin{figure}
    \centering
    \includegraphics[width=0.8\textwidth]{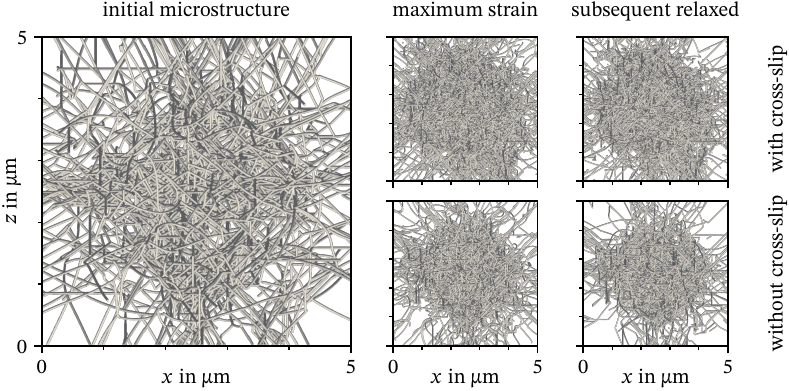}
    \captionbelow{ Examples of the dislocation microstructures observed for one
      realization viewed along the tensile axis.  The initial structure (left)
      evolves differently with (top row) or without (bottom row) cross-slip.
      Evolved microstructures are shown at the peak strain just before the
      relaxation (middle column) and at the end of the relaxation (right
      column).}
    \label{fig:discrete-dislocation-structures}
\end{figure}

The initial dislocation microstructure consists of dislocation loops with randomly selected radii between \SIrange{2}{8}{\micro\meter} such that their centers are in a volume that is four times the size of the simulation box.
This way, the simulation box comprises whole loops and segments that end at its surfaces.
Dislocations are uniformly drawn from all 12 possible slip systems.
Subsequently, the system is allowed to relax, i.e., we evolve it in time without applied external load until an equilibrium dislocation structure is reached.
This process is tuned such that after the relaxation, the total dislocation density of each initial dislocation microstructure is close to \SI{1.15e13}{1/\square\meter}.
We generated \num{10} realizations.
An example for the initial dislocation microstructure is shown in the left column of \cref{fig:discrete-dislocation-structures};
further details can also be found in \autocite{steinberger_thesis}.
We then perform tensile tests with these initial dislocation microstructures along the $y$-axis twice for each realization, once with cross-slip and once without.
Displacements are prescribed at the top in positive $y$ direction with a strain rate of \SI{5000}{\per\second}. The bottom surface is fixed ($u=0$).
Snapshots of the dislocation microstructure are saved periodically during loading and subsequent unloading:
At a strain of about \SI{0.6}{\percent}, we stop the tensile test and allow the dislocation microstructure to relax without external load.
Examples for dislocation microstructures at maximum strain and after relaxation are shown in the center and right column of \cref{fig:discrete-dislocation-structures}, respectively. The top row shows microstructures with cross-slip allowed, the bottom without.

\subsection{Discrete-to-continuous method}

Within the \ab*{discrete-to-continuous} method~\autocites{Sandfeld_2015_c}{Steinberger_2016_b}, we treat each dislocation as a parameterized directed curve $\mathcal{C}(t)$. $t \in [a, b]$ in $\mathcal{C}(t)$ denotes the parametrization where $a$ and $b$ are the start and end positions of the dislocations.
In addition to the spatial location of all points of the curves in space, we associate each curve with the Burgers vector of the dislocation that it represents. Then, treating dislocations as curves allows us to conveniently compute quantities such as the tangent vector
\begin{equation}
    \hat{\xi}(t)
    =
    \frac{
        \mathcal{C}^{\prime}(t)
    }{
        \norm[0]{\mathcal{C}^{\prime}(t)}
    }
\end{equation}
and the unsigned curvature
\begin{equation}
    k(t)
    =
    \frac{
        \sqrt{
            \norm[0]{\mathcal{C}^{\prime}(t)}^{2} \norm[0]{\mathcal{C}^{\prime\prime}(t)}^2
            -
            \del[1]{
                \mathcal{C}^{\prime}(t) \cdot \mathcal{C}^{\prime\prime}(t)
            }^{2}
        }
    }{
        \norm[0]{\mathcal{C}^{\prime}}^{3}
    },
\end{equation}
where $\mathcal{C}^{\prime}(t)$ and $\mathcal{C}^{\prime\prime}(t)$ denote the first and the second derivative of $\mathcal{C}(t)$ with respect to $t$.
To compute dislocation density fields $\bullet$, we first discretize the domain $\Omega$ into $n(\mathcal{V})$ subvolumes $\Omega_{i}$.
Within a subvolume, we may compute the quantity of interest via
\begin{equation}
    \bullet_{\Omega_{i}}
    =
    \frac{1}{V_{\Omega_{i}}}
    \sum \limits_{\mathcal{C}}
    \int \limits_{\mathcal{C} \in \Omega_{i}}
    f_{\mathcal{C}}^{\bullet}(t) \norm[0]{\mathcal{C}^{\prime}(t)}
    \dif t,
\end{equation}
where $V_{\Omega_{i}}$ denotes the volume of the subdomain $\Omega_{i}$, and $f_{\mathcal{C}}(t)$ denotes a function whose expression depends on the continuum field ${\bullet}$ to be computed.
For example, if we use $f_{\mathcal{C}}(t) = 1$, we obtain the total dislocation density $\rho^{(0)}$, and with $f_{\mathcal{C}}(t) = k_{\mathcal{C}}(t)$ we obtain the curvature density, denoted by $q^{(0)}$.
For more details, see~\autocite{Steinberger_2019_a}.

\subsection{Averaging and comparing dislocation microstructures}
Discretizing dislocation microstructures within equal domains using a fixed discretization scheme allows us to average and compare dislocation microstructures quantitatively.

%
%
The comparison of dislocation microstructures is carried out on two levels: field values in subvolumes and whole simulations.
First, we introduce a measure of deviation between the field values in a subvolume.
For the comparison of scalar fields we use the absolute difference:
\begin{equation}
    D^{\Omega_{i}}(\bullet_{\Omega_{i}}, \circ_{\Omega_{i}})
    =
    \abs[0]{
        \circ_{\Omega_{i}} - \bullet_{\Omega_{i}}
    }
\end{equation}
where the fields from two separate dislocation microstructures ($\bullet$ and $\circ$),  within a subvolume are respresented by $\bullet_{\Omega_{i}}$ and $\circ_{\Omega_{i}}$.
For the comparison of two whole simulations (domains), we use the weighted average absolute difference
\begin{equation}
	\label{eq:weighted-absolute-difference}
    D^{\Omega}(\bullet, \circ)
    =
    \frac{1}{V_{\Omega}}
    \sum_{i = 1}^{n(\mathcal{V})}
        D^{\Omega_{\mathrm{i}}}(\bullet_{\Omega_{i}}, \circ_{\Omega_{i}})
    \,V_{\Omega_{i}},
\end{equation}
where $V_{\Omega}$ denotes the volume of the domain, and $V_{\Omega_{i}}$ is the volume of a subvolume as introduced earlier.


%
%
A set of fields, $\bullet$, (either total dislocation density or curvature in this study) of dislocation microstructures $\mathcal{S} = \{ \bullet^{1}, \bullet^{2}, \dotsc{}, \bullet^{n(\mathcal{S})} \}$ is extracted using the \ab*{discrete-to-continuous} method.
The average value of a field within a given subvolume $\Omega_{i}$ is then computed via
\begin{equation}
    \label{eq:ensemble-average}
    \langle \bullet \rangle_{\Omega_{i}}^{\mathcal{S}}
    =
    \frac{1}{n(\mathcal{S})}
    \sum \limits_{j=1}^{n(\mathcal{S})}
    \bullet_{\Omega_{i}}^{j}.
\end{equation}

%
%
The \ab*{mean absolute deviation} is used as a measure of how different the dislocation microstructures within a set are to each other in terms of a selected field variable:
\begin{equation}
    \text{\as{mean absolute deviation}}
    =
    \frac{1}{n(\mathcal{S})}
    \sum_{j}^{n(\mathcal{S})}
    \abs[1]{
        D^{\Omega}
        \del[1]{
            \bullet^{j}, \langle \bullet \rangle^{\mathcal{S}}
        }
    }.
\end{equation}
Lower values indicate higher similarity within a set. But a comparison of values between different sets of dislocation microstructures is not meaningful as their average values of, e.g., the total dislocation density, might be very different.
To enable a comparison across microstructure, we first compute the domain average
\begin{equation}
	\label{eq:domain-average}
    \langle \bullet \rangle^{\Omega}
    =
    \frac{1}{V_{\Omega}}
    \sum_{i = 1}^{n(\mathcal{V})}
    \bullet_{\Omega_{i}}
    V_{\Omega_{i}},
\end{equation}
of our field quantity of interest and use it to compute the unitless \ab*{coefficient of variation} of the \ab*{mean absolute deviation}
\begin{equation}
    \text{\as{coefficient of variation}}\del[1]{\text{\as{mean absolute deviation}}\del[0]{\mathcal{S}}}
    =
    \frac{
        \text{\as{mean absolute deviation}}\del[0]{\mathcal{S}}
    }{
        \bigl\langle \langle \bullet \rangle^{\mathcal{S}}_{\Omega_{i}} \bigr\rangle^{\Omega}
    }.
\end{equation}

This measure of dispersion around the average can then be used to compare the deviation of the microstructures of a set with the one of another set.

\section{Results}
%
%

\begin{figure}
	\centering
	\begin{subfigure}[t]{0.4\textwidth}
		\includegraphics[width=\textwidth]{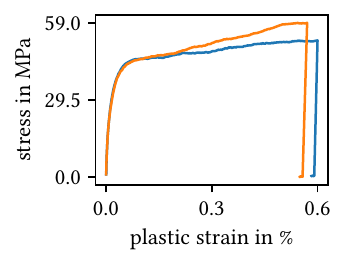}
		\caption{Average stress along the tensile direction over plastic strain in the same direction for the two sets of simulations with and without cross-slip.}
		\label{fig:eps_pl-vs-sig}
	\end{subfigure}\hspace{1cm}
	\begin{subfigure}[t]{0.52\textwidth}
		\includegraphics[width=\textwidth]{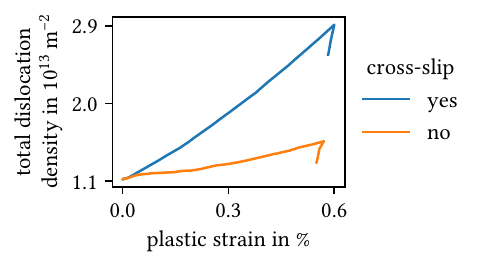}
		\caption{Average total dislocation density over plastic strain in the same direction for the two sets of simulations with and without cross-slip.}
		\label{fig:eps_pl-vs-rho}
	\end{subfigure}
	\caption{Average stress and total dislocation density evaluation of the simulations over the plastic strain.}
	\label{fig:eps_pl}
\end{figure}

The averaged tensile stress and averaged dislocation density over the plastic strain in the $y$-direction for both
simulation sets with and without cross-slip is shown in \cref{fig:eps_pl}. The evaluation of the average stress in \cref{fig:eps_pl-vs-sig} can be divided into 3 phases. In the first phase, the average stress has the same steep increase for both cross-slip and without cross-slip cases. 
It is mainly elastic deformation with the first indication of microplasticity. 
In the second phase, the average stress increases linearly for both cases, but the slope is higher for without cross-slip. The mismatch of the maximum plastic strain for two cases is because of the adaptive time steps used in the simulations. The third phase is the relaxation phase where the stress drops to zero due to the removal of load and followed by the small decrease in the plastic strain due to the Bauschinger effect. These phases can also be tracked in \cref{fig:eps_pl-vs-rho}: First, both cases show similar increase in the average total density, then higher increase in the case of cross-slip is observed. Finally, the density decreases in both of the cases during the relaxation phase.

\begin{figure}[htp]
	\centering
	\begin{subfigure}[t]{0.4\textwidth}
		\includegraphics[width=\textwidth]{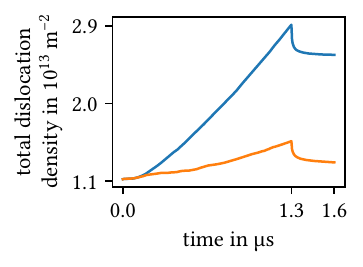}
		\caption{Average total dislocation density along the tensile direction over time for the two sets of simulations with and without cross-slip.}
		\label{fig:time-rho-all}
	\end{subfigure}\hspace{0.5cm}
	\begin{subfigure}[t]{0.52\textwidth}
		\includegraphics[width=\textwidth]{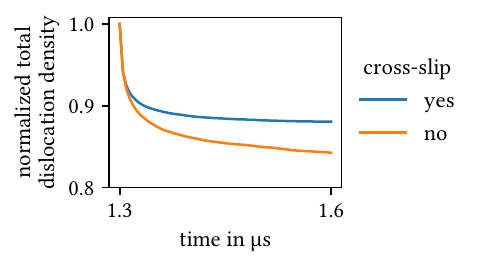}
		\caption{Average normalized total dislocation density over time since the maximum strain for the two sets of simulations with and without cross-slip.}
		\label{fig:time-rho-relax}
	\end{subfigure}
	\caption{Total dislocation density evaluation of the simulations over time.}
	\label{fig:time-rho}
\end{figure}

%
The average evolution of the total dislocation density $\rho_{\Omega}^{(0)}$ of the whole domains $\Omega$ for both with and without cross-slip is shown in \cref{fig:time-rho-all}. During the initial \SI{0.2}{\micro\second}, all microstructures show the same small increase in total dislocation density irrespective of cross-slip or not.
After this initial stage, the total dislocation density increases more rapidly with cross-slip during loading than without cross-slip. Upon releasing the external load after \SI{1.3}{\micro\second}, the dislocation density initially drops sharply and then continues to decrease but at a much slower rate.  

For a simpler comparison of the decrease in the density during relaxation, we show the average total dislocation density normalized by the maximum density as a function of time starting with the time of the external load release in  \cref{fig:time-rho-relax}. The initial relative decrease in the average total dislocation density is comparable for all simulations. However, after about \SI{0.1}{\micro\second} we observe that no cross-slip leads to a larger decrease in the relative total dislocation density compared to with cross-slip.
While the former drops by about \SI{15}{\percent}, the latter only decreases by little more than \SI{10}{\percent}.

\minisec{Features of the dislocation microstructures}

In the \emph{initial state}, we observe the highest concentration of dislocations in the center of the $xz$-plane with a noticeable drop-off within about \SI{1}{\micro\meter} distance to the surface.
Dislocations align approximately perpendicular to the surface.

At maximum strain, the total density of the microstructure strongly depends on cross-slip.
In case without, we observe an increase of the density in the center portion of the $xz$-plane.
The depletion close to the free surface is similar to the initial microstructure.
With cross-slip we see an even stronger increase in the density in the center of the $xy$-plane as well as
a higher dislocation density closer to the free surfaces.

Upon relaxation we observe a depletion of dislocation near the free surfaces in
both cases.
However, with cross-slip a denser dislocation structure can be observed in surface-near regions compared to without cross-slip.

\section{Discussion}

\subsection{Influence of cross-slip on the dislocation microstructure}

The prominent differences between dislocation microstructures with and without cross-slip are higher total dislocation densities and dislocations present closer to the open surfaces.
While the former is evident from \cref{fig:time-rho-all} and in line with similar numerical experiments~\autocites{Motz_2009_a}{Zhou_2010_a}{Hussein_2015_a}, we have only shown one realization per set in \cref{fig:discrete-dislocation-structures}.
By performing ensemble averages of each set via \cref{eq:ensemble-average}, we investigate whether this is a general observation of considering cross-slip or merely an outlier of the shown realizations.
The average total dislocation density for each set of dislocation microstructures further averaged along the $y$-direction is shown in \cref{fig:total-dislocation-density-averaged-along-y}.

\begin{figure}[htp!]
    \centering
    \includegraphics[width=0.8\linewidth]{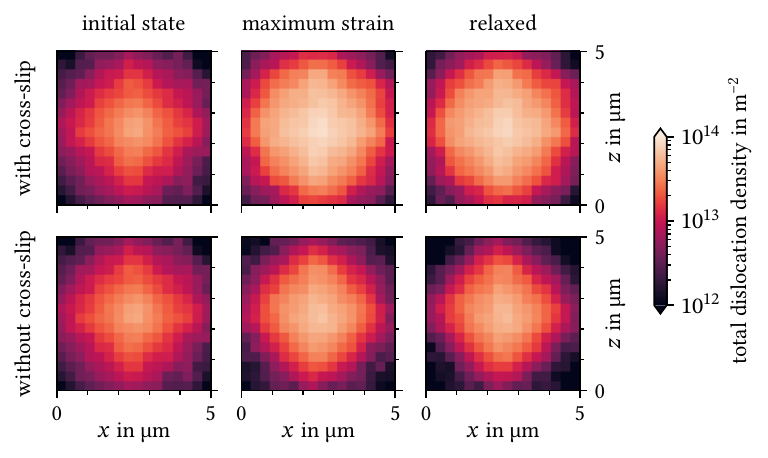}
    \captionbelow{%
        Average total dislocation density per dislocation microstructure set further averaged along the $y$-direction, i.e., the viewing direction of \cref{fig:discrete-dislocation-structures} and this figure.
        The spatial discretization is \num{16} voxels along each direction.
    }
    \label{fig:total-dislocation-density-averaged-along-y}
\end{figure}

We can see that the higher probability of dislocations being present closer to the surface is a feature in the set of simulations with cross-slip.
The reason for this is the two-fold nature of cross-slip with regards to dislocation motion.
On the one hand, cross-slip enables dislocations to move on a second glide plane in three dimensions instead of being confined to their glide plane.
On the other hand, the motion of the part of the dislocation where cross-slip occurred is confined to the intersection line of the two slip planes on which it took place. Therefore, the dislocation is restricted by the intersection of primary and cross-slip plane and only able to move in one dimension at the cross-slip site.  These two contributions result in more space for dislocations to evolve but limited mobility, hence the dislocation density at the surface is stabilized. This is beyond a simple scaling with the total dislocation density as discussed later by evaluating the relative dislocation density evaluation in the regions close to the surface and within the center region. 

As the dislocations move, they act as obstacles for other dislocations and may also form junctions that restrict the dislocation motion. 
To overcome these obstacles, dislocations can cross-slip onto other slip planes.
And while some segments of the dislocation might be able to move on its new slip plane, the segment connected to the primary dislocation is not be able to overcome other obstacles in this manner due to its reduced degrees of freedom.
This effectively means that cross-slip both adds a degree of freedom to the motion of dislocations overall at the expense of limiting the motion of parts of the dislocation where the cross-slip originated from and thereby also stabilizing the structure.
In combination, more space is available for dislocations to move with potentially more obstacles to get stuck at, even in the presence of attractive image forces due to free surfaces. Therefore, cross-slip stabilizes dislocation densities close to surfaces.

In combination with \cref{fig:time-rho-relax}, we conclude that the stabilizing characteristics of cross-slip affect the subsequent relaxation because the average relative decrease in the total dislocation density is smaller for realizations with cross-slip.

\begin{figure}[htp!]
	\centering
	\includegraphics[width=0.6\linewidth]{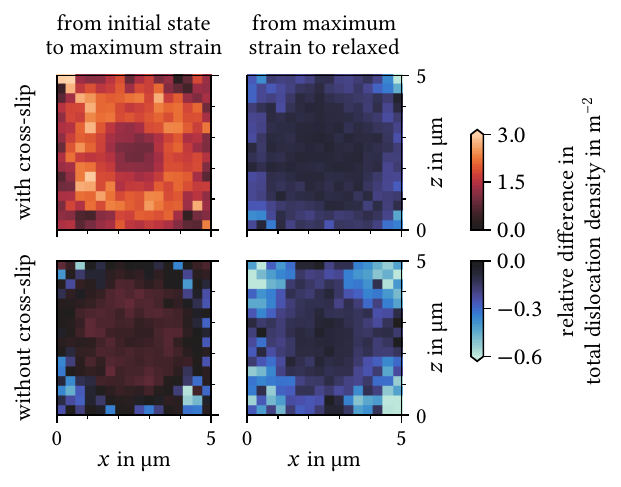}
	\captionbelow{Average relative change in the total dislocation density per
		dislocation microstructure set further averaged along the $y$-direction,
		i.e., the viewing direction of \cref{fig:discrete-dislocation-structures}
		and this figure.  Changes between the dislocation microstructures from
		their initial state to the maximum strain state and from the latter to
		their relaxed state are depicted in the left and right column,
		respectively. The spatial discretization is \num{16} voxels along
		each direction.}
	\label{fig:total-dislocation-density-differences_averaged-along-y}
\end{figure}

We now address how the change in density is spatially distributed within the specimens.
\cref{fig:total-dislocation-density-differences_averaged-along-y} shows the average relative change in the total dislocation density for the two sets of realization between the initial state and the maximum strain state, as well as between the maximum strain state and the subsequent relaxed state.

With cross-slip, the relative increase is higher closer to the free surfaces.
Without cross-slip, we see an increase in the center and the regions close to the open surfaces of the sample while there is decrease in the total dislocation density at the open surfaces. 

During unloading, the largest relative decreases in the total dislocation density are observed close to the surfaces.
With cross-slip, the difference between the decrease close to the surfaces compared to the one in the center of the sample is smaller compared to realizations without cross-slip.
Hence, we conclude that this further confirms the stabilizing effect of cross-slip.

\subsection{Similarity of dislocation microstructures across sets}
\begin{figure}[htp]
    \centering
    \includegraphics[width=0.75\textwidth]{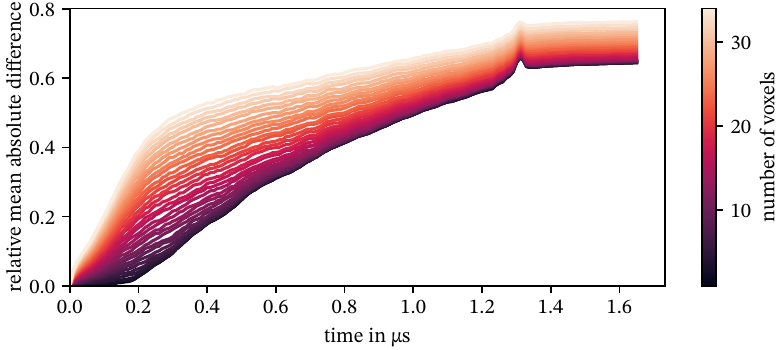}
    \captionbelow{Relative mean absolute difference of the total dislocation
      densities between the microstructures with and without cross-slip.  Colors
      indicate the numbers of voxels that are used to discretize the specimen for the
      computation of the continuum fields}
    \label{fig:coefficient-of-variation-between-sets}
\end{figure}

We show the relative mean absolute difference of two dislocation microstructure sets for the total density over time for different spatial discretizations in \cref{fig:coefficient-of-variation-between-sets}. The data points in the lines in \cref{fig:coefficient-of-variation-between-sets} are calculated as follows:

\begin{equation}
	\Delta\rho^{(0)}(t) = 
	\frac{
		\left\langle 
			\left\langle
				\abs[0]{		
					\rho^{(0)}_{\mathcal{S}_{0,j}}(t) - \rho^{(0)}_{\mathcal{S}_{1,j}}(t)
			}	
			\right\rangle_{\Omega_{i}}^{\mathcal{S}} 
		\right\rangle^{\Omega}
	}
	{
		\left\langle 
			\left\langle 
				\rho^{(0)}_{\mathcal{S}_{0}, \mathcal{S}_{1}}(t)
			\right\rangle_{\Omega_{i}}^{\mathcal{S}} 
		\right\rangle^{\Omega}
	}
\end{equation}
where $\langle \bullet \rangle_{\Omega_{i}}^{\mathcal{S}}$ and $\langle \bullet \rangle^{\Omega}$ are defined in \cref{eq:ensemble-average} and \cref{eq:domain-average}, respectively. $\mathcal{S}_{0,j}$ and $\mathcal{S}_{1,j}$ indicate the specimen pairs, which have the same initial microstructures, from the set of samples without cross-slip $(\mathcal{S}_{0})$ and with cross-slip $(\mathcal{S}_{1})$. The denominator calculates the average value by using the density values of the subvolumes of all specimens to normalize the difference.  Finally, $\Delta\rho^{(0)}(t)$ is the relative mean absolute difference of two dislocation microstructure sets for the total density at a given time step and discretization resolution.

Initially, there is no difference as we start from the exact same microstructures.
Upon loading and deformation of the sample we observe an increase in the relative mean absolute difference that strongly depends on the spatial discretization.
Higher resolutions are more suitable to show increasing differences between the structures with and without cross-slip as microstructures discretized with higher resolutions contains more information about the actual location of the dislocations.
The largest spread of the difference is observed at about \SI{0.2}{\micro\second}, which is around the end of the elastic regime. Afterwards, the magnitude of the relative mean absolute difference between the spatial discretizations decreases again.
From this we may conclude that the initial differences observed are primarily on a rather short length scale and therefore only seen for high resolutions.
As the plastic strain accumulates, the changes in the topology of the dislocation microstructure cover a larger length scale and we observe an increase in the difference for coarse spatial resolutions as well.

At about \SI{1.3}{\micro\second} a spike occurs.
This coincides with the unloading for realizations without cross-slip.
While these realizations exhibit a rather severe and fast change in the dislocation microstructure, the realizations with cross-slip show a more stable behavior.

\subsection{Similarity of dislocation microstructures within each set}
\label{subsec:similarity}
\begin{figure}[htp!]
    \centering
    \includegraphics{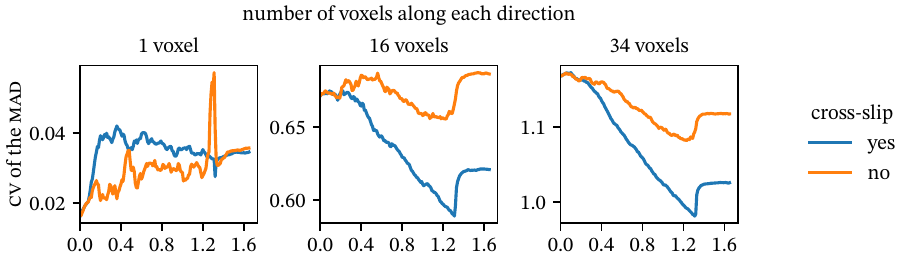}
    \captionbelow{ \Al{coefficient of variation} of the \al{mean absolute
        deviation} of total dislocation density fields of each dislocation microstructure set over time for a
      spatial discretization of \numlist{1;16;34} voxels along each direction.
    }
    \label{fig:coefficient-of-variation-over-time-slices}
\end{figure}

We know that differences may manifest on different length scales from our previous discussion. Thus, to study the similarity of realizations \emph{within each set}, we show the \ab*{coefficient of variation} of the \ab*{mean absolute deviation} of the total dislocation densities for different spatial discretizations in
\cref{fig:coefficient-of-variation-over-time-slices}.
The first thing to note is that the CV values depend strongly on the discretization.
Discretizations with higher spatial resolution show larger dissimilarity overall.
This stems from coarser discretizations averaging over many more microstructure features with less sensitivity for their actual position in space. 
In contrast, finer discretized microstructures are more sensitive to the position of the dislocation and therefore they are similar to each other when the dislocations do not match up closely between different realizations.

Irrespective of the discretization, the dislocation microstructure sets exhibit the same \gls{coefficient of variation} of the \gls{mean absolute deviation} in the beginning.
This is due to the fact that realizations are using the exact same initial structures and require some ``incubation'' time of mainly elastic deformation until more and more dislocations start to move.
After about \SI{0.1}{\micro\second}, the samples' behavior starts to differ with the onset of plastic deformation.
For \num{1} voxel along each direction, the realizations without cross-slip are more similar to each other than the ones with cross-slip.
The exception at around \SI{1.3}{\micro\second} comes from the staggered onset of the relaxation sequence for the realizations without cross-slip.
For higher spatial resolutions, this trend is reversed and the dislocation microstructures with cross-slip are more similar to each other than the ones without.

\begin{figure}[htp]
    \centering
    \includegraphics{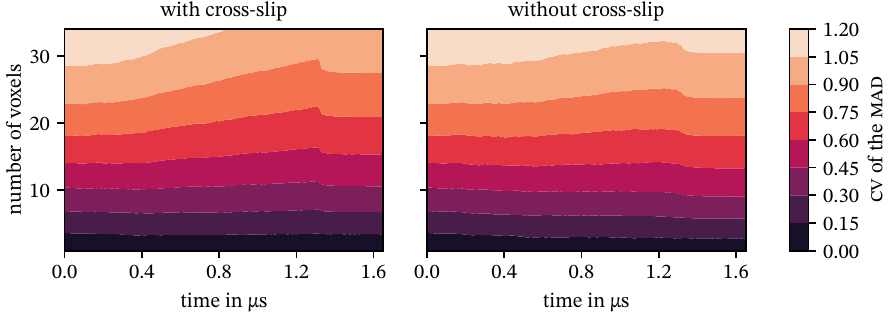}
    \captionbelow{%
        CV of the \al{mean absolute deviation} of total dislocation density fields of each set of cross-slip considerations over time for different resolutions of the spatial discretization.
    }
    \label{fig:coefficient-of-variation-over-time}
\end{figure}

The evolution of the \ab*{coefficient of variation} of the \ab*{mean absolute deviation} for discretizations using \numrange{1}{34} voxels along each direction is shown in \cref{fig:coefficient-of-variation-over-time}.

Darker colors indicate higher similarity between the dislocation microstructures of a set.
We observe the previously mentioned trend that a higher resolution results in a larger dissimilarity.

During the tensile test, we notice a tendency of the microstructures becoming more similar to each other as the tensile tests progress, particularly for discretizations using more than \num{15} voxels along each direction.
This trend is more pronounced for simulations with cross-slip.
We conclude that the spatial arrangement of the dislocation microstructure becomes increasingly more similar over the course of the tensile test.
Assuming that there are relatively stable and/or favorable dislocation configurations that form during loading, the inclusion of cross-slip as a degree of freedom explains why the similarity between realizations with cross-slip increases more rapidly than that of the realizations without cross-slip.

\subsection{Probability density functions for total density and line curvature}
\begin{figure}
	\centering
	\includegraphics[width=\linewidth]{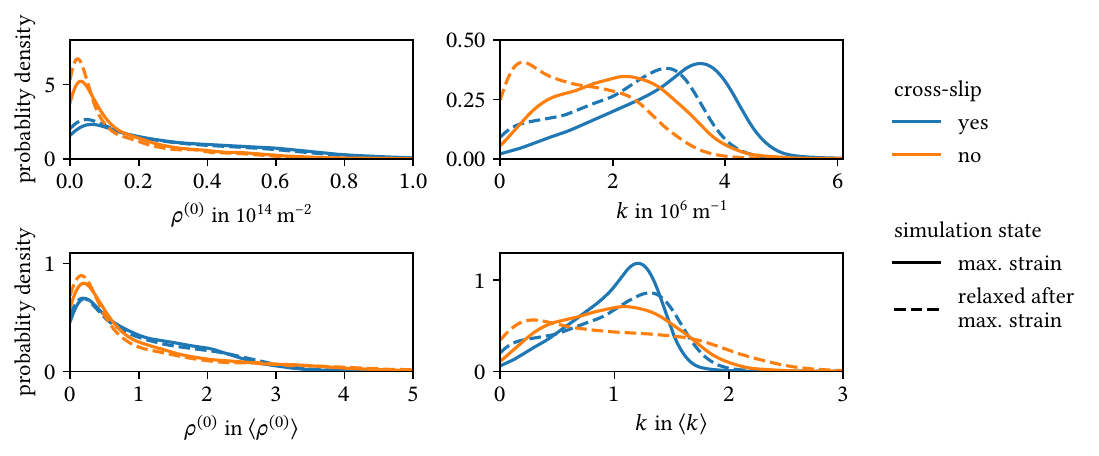}
	\caption{Probability density functions of the total dislocation density and the
		curvature observed for the averaged steady state configurations of all evolved initial
		dislocation microstructures discretized by 16 voxels in each axis. The top row is based on the absolute values, the bottom
		row is normalized with the sample mean of each set.}
	\label{fig:pdfrhocurve}
\end{figure}

While the total density (where the average is taken by assuming only one voxel) is a field variable that is commonly used for analyzing dislocation data, the lines' curvature are often not assessed for analyzes.
With the \ab*{discrete-to-continuous} framework this can be easily done: the mean curvature can be directly computed from the curvature density and the total density, $k=q^{(0)}/\rho^{(0)}$. \Cref{fig:pdfrhocurve} shows the probability density functions of the total dislocation density,  $\rho^{(0)}$, and the curvature, $k$. The quantities are averaged over multiple realizations, then their probability density functions are calculated via kernel density estimation.

When cross-slip is activated, we observe more even distribution of the total dislocation density as the probability of low densities increases significantly in the case of without cross-slip. However, the probabilities of the normalized total dislocation densities from the cross-slip and without cross-slip cases become closer in terms of their values. In the distributions of the curvature, the difference between the cases of with and without cross-slip is more pronounced compared to the total dislocation density distributions. Moreover, when the curvature values are normalized, the difference between two cross-slip cases becomes even larger in terms of the values of the probabilities which in contrasts with what is observed when the total dislocation density is normalized. For the simulations with cross-slip, a pronounced a high probability density of curvature values of about $1.2 \text{ } \langle k\rangle$ becomes clearly visible.
This seems to be an aspect to be considered for initial field values for \ab*{discrete dislocation dynamics} or \ab*{continuum dislocation dynamics} simulations. Furthermore, it might be a way of ``testing'' if the simulation was run with cross-slip enabled or not.

\subsection{Effects of discretization on total density and line curvature distributions}

\begin{figure}[htp!]
	\centering
	\includegraphics[width=1\linewidth]{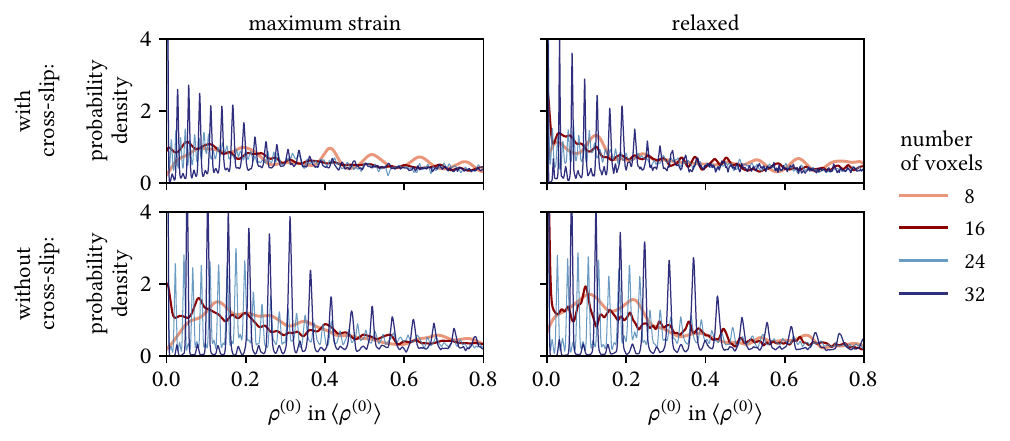}
	\caption{Probability density distributions of the normalized total density for the discretizations by 8, 16, 24, and 32 voxels along each direction. The first and second row show the results from simulations with cross-slip and without cross-slip, respectively. The first and second columns show the results from simulations that are at the maximum strain state and the subsequent relaxed state, respectively.}
	\label{fig:pdfrhomulti}
\end{figure}

\begin{figure}[htp!]
	\centering
	\includegraphics[width=1\linewidth]{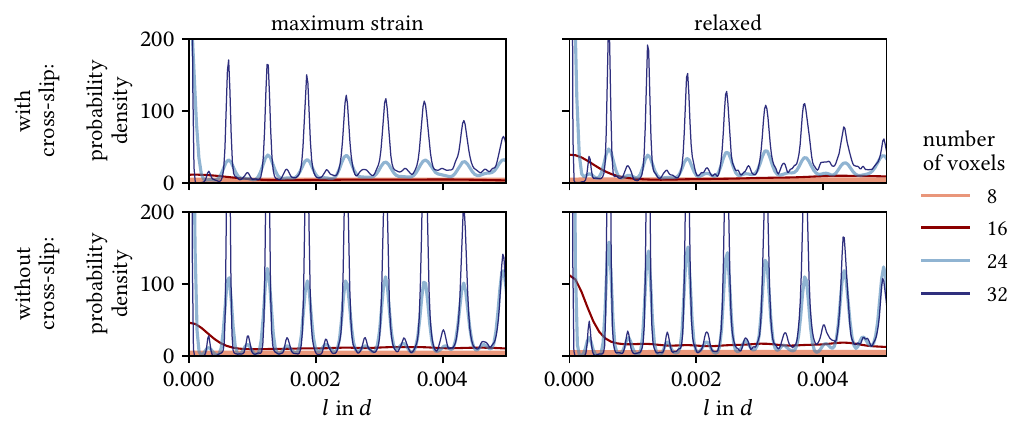}
	\caption{Probability density distributions of the dislocation length in the voxels for the discretizations by 8, 16, 24, and 32 voxels along each direction. $l$ is the total dislocation length in a voxel, and $d=\SI{5.0}{\micro\meter}$ is the specimen edge length, which is used for normalization of the $l$ values. The first and second row show the results from simulations with cross-slip and without cross-slip, respectively. The first and second columns show the results from simulations that are at the maximum strain state and the subsequent relaxed state, respectively.}
	\label{fig:pdflengthmulti}
\end{figure}

We further investigate how the distribution of the total density and line curvature fields over the domain changes for different discretizations.
In \cref{fig:pdfrhomulti}, the probability density functions of the normalized total dislocation density for different  discretizations (8, 16, 24 and 32 voxels along each axis) are shown. The comparison of the distributions for the discretizations with 16, 32, 40, 64 voxels along each axis is provided in \cref{fig:pdfrhomulti_more}. 
Probability densities are calculated by kernel density estimations, and we used the improved Sheather-Jones algorithm~\autocite{botev2010kernel} to decide on the optimum bandwidth values to represent the distributions accurately.
Through this, we plotted distributions as if they were represented by histograms with a very high number of small bins. 

\begin{figure}
	\centering
	\begin{subfigure}[t]{0.6\textwidth}
		\centering
		\includegraphics[width=\linewidth]{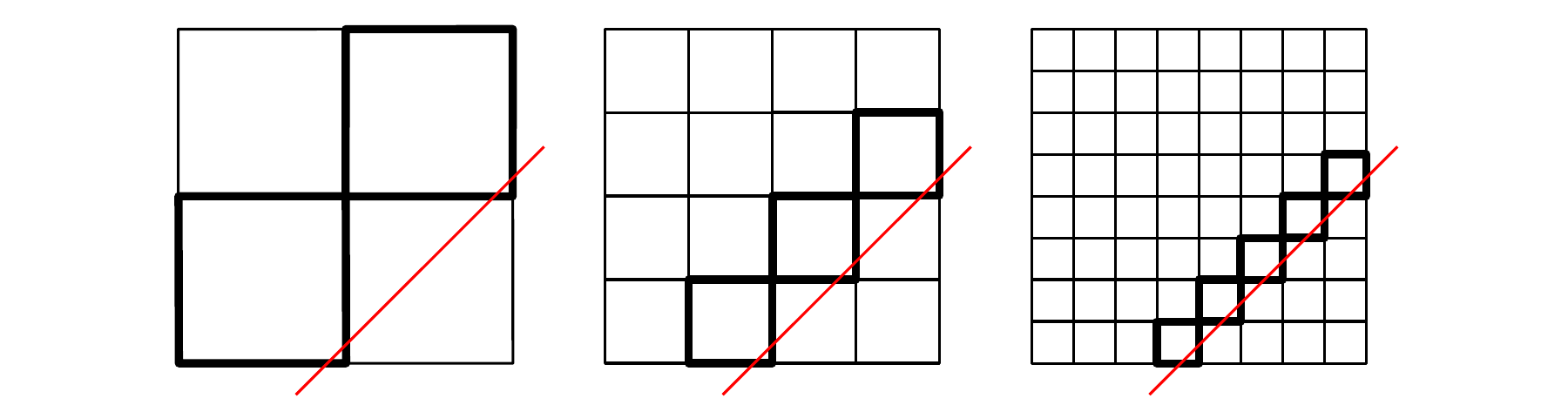}
		\caption{Demonstration of the increase in the number of pixels that contain the same length with an increasing resolution of discretization. The pixels plotted with thicker edges contain the same line length in all discretizations.}
		\label{fig:discr-singlegray}
	\end{subfigure}\hspace{1cm}
	\begin{subfigure}[t]{.2\textwidth}
		\centering
		\includegraphics[width=\linewidth]{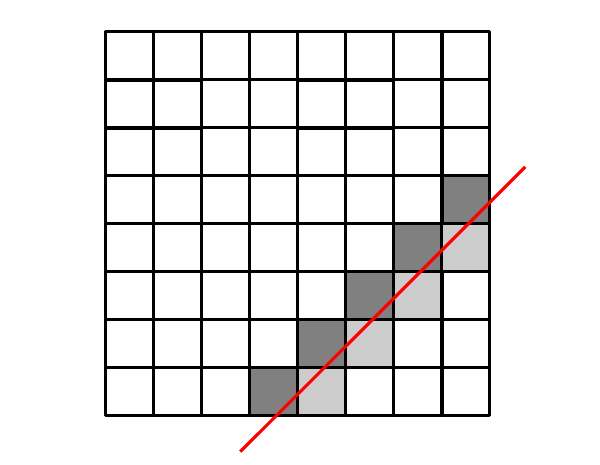}	
		\caption{Representation of the pixels that contain ``two particular'' values of the line length (light gray and dark gray).}
		\label{fig:discr-doublegray}
	\end{subfigure}
	\caption{Discretization of a straight line in two-dimensions.}
	\label{fig:discr}
\end{figure}

The first notable observation is the occurrence of equally spaced fluctuations or oscillations in the distributions for each type of the specimen (with and without cross-slip) and state of the simulation (maximum strain and relaxed) when specimens discretized with higher resolutions.
The formation of the peaks in the total density distributions are due to the geometrical reasons rather than the physical dislocation behavior or the crystallographic orientation of the specimen: as we increase the resolution of the discretization, we are approaching a discrete representation of the dislocation lines.
Considering this together with a large number of straight, diagonal lines in the simulation volume, the result is a high number of voxels that contain the same total line length.
We also notice that the spacing between the peak points are inversely proportional to the volume of the voxels for each case in \cref{fig:pdfrhomulti}.
This means that the distribution of the total dislocation length in the voxels has peak points at the same values for all discretizations as shown in \cref{fig:pdflengthmulti}. (See \cref{fig:pdflengthmulti_more} for the distributions in the case of discretizations with 16, 32, 40, 64 voxels along each axis.)
Geometrical reasons behind the formation of the peaks are simply demonstrated in two-dimensional geometries in \cref{fig:discr}.
In \cref{fig:discr-singlegray}, as the discretization becomes finer, the number of the subvolumes that have the same length of the line (the subvolumes with thicker edges) increases.
This leads to the formation of local peak values in the distributions plots.
Apart from this, the discretization of one line results in paired subvolumes with respect to the contained length (cf. the highlighted pixels by light and dark gray in the \cref{fig:discr-doublegray}).
This leads to the formation of multiple peak points in the distributions.

For the same level of discretization, more pronounced fluctuations are observed without cross-slip.
In other words, the total density fields of the specimens with cross-slip are less sensitive to the discretization as compared to the ones without cross-slip.
The lower sensitivity to the discretization is desirable for continuous field data since we can access more information by increasing the resolution before the discrete information starts to dominate the distribution.
The difference in sensitivity between the cases with and without cross-slip is another result of the stabilizing effect of cross-slip, which was already mentioned in the previous analyzes.
For both specimens, we do not observe a significant difference between the distributions of the maximum strain state and the relaxed state. 

In addition, observing strong fluctuations at finer resolutions than 16 voxels for the case without cross-slip and 24 for the case including cross-slip is in line with the conclusions in~\autocite{Steinberger_2019_a} as the authors proposed the average dislocation spacing as a lower limit for the voxel size due to the physical considerations. The reason is that for both numbers of voxels the voxel sizes are smaller than the average dislocation spacing. Although we capture more differences in very high resolutions between microstructures, domains discretized with a voxel size larger than average dislocation spacing are not greatly affected by geometrically induced distortions in the distributions.

\begin{figure}[htp!]
	\centering
	\includegraphics[width=0.9\linewidth]{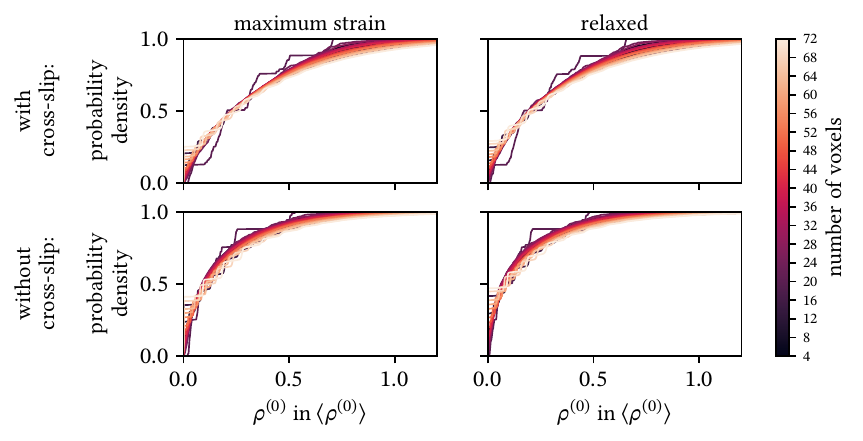}
	\caption{Cumulative density distributions of the normalized total density for the discretizations by 18 different number of voxels along each direction. The first and second row show the results from simulations with cross-slip and without cross-slip, respectively. The first and second columns show the results from simulations that are at the maximum strain state and the subsequent relaxed state, respectively.}
	\label{fig:cdfrhomulti}
\end{figure}

\begin{figure}[htp!]
	\centering
	\includegraphics[width=0.9\linewidth]{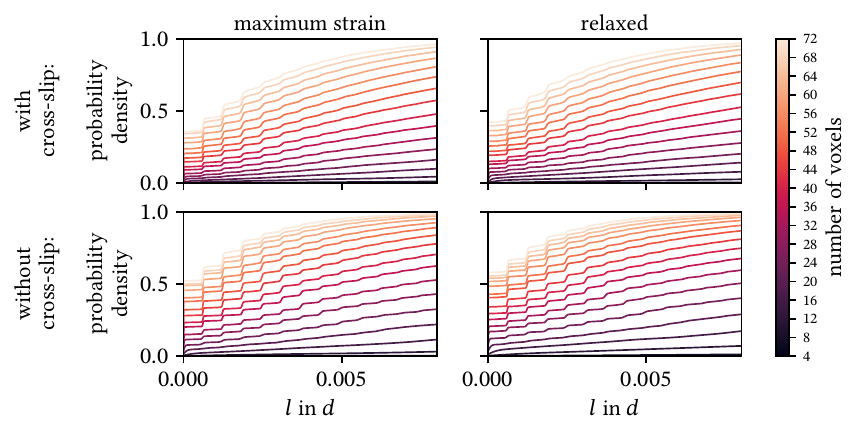}
	\caption{Cumulative density distributions of the dislocation length in the voxels for the discretizations by 18 different number of voxels along each direction. $l$ is the total dislocation length in a voxel, and $d=\SI{5.0}{\micro\meter}$ is the specimen edge length, which is used for normalization of the $l$ values. The first and second row show the results from simulations with cross-slip and without cross-slip, respectively. The first and second columns show the results from simulations that are at the maximum strain state and the subsequent relaxed state, respectively.}
	\label{fig:cdflengthmulti}
\end{figure}

To show the effect of the discretization based on a larger range of resolutions, we  calculated cumulative density functions to improve readability.
In \cref{fig:cdfrhomulti}, the cumulative distributions are shown for 18 different discretizations.
The step-function-like shape of the distributions corresponds to the fluctuations in probability density functions, and they are less pronounced with cross-slip.
In \cref{fig:cdflengthmulti}, we further show cumulative density distributions for the line lengths. There, fluctuations are observed at the same values of the line length for each discretization. This is consistent with the observation of the spikes which occur at the same values at each discretization in the probability density distributions of normalized line length values.

\begin{figure}
	\centering
	\includegraphics[width=0.9\linewidth]{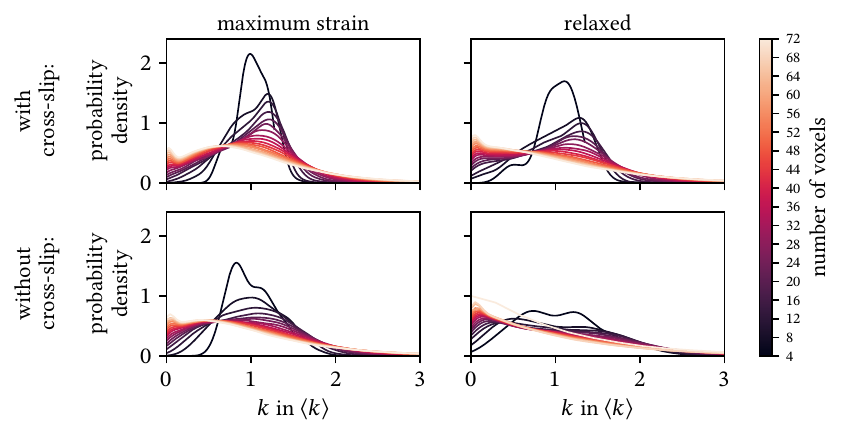}
	\caption{Probability density distributions of the normalized curvature for the discretizations by 18 different number of voxels along each direction. The first and second row show the results from simulations with cross-slip and without cross-slip, respectively. The first and second columns show the results from simulations that are at the maximum strain state and the subsequent relaxed state, respectively.}
	\label{fig:pdfkmulti}
\end{figure}

We repeat the previous analysis for the normalized line curvatures, which are obtained by averaging over multiple realizations, and the comparison of these distributions for 18 different discretization levels are shown in \cref{fig:pdfkmulti}.
We can see the effect of the cross-slip on the curvature distributions on the relaxed states of the specimens: the peak is closer to the median of the data sample in the case of cross-slip. The effect on the shapes of the curves is demonstrated in \cref{fig:skewness} by the skewness values of the distributions.

Hence, the deduction, from section \cref{subsec:similarity}, that cross-slip has an effect on the curvature distribution shape is also true for different discretizations.
Similar to our observations for the total density distributions, we do not observe a significant change in the curvature distributions after the load is removed at the maximum strain state both for the specimen with cross-slip and without cross-slip. 

\section{Conclusion}

We studied the impact of cross-slip on the evolution of dislocation microstructures, which are obtained by \ab*{discrete dislocation dynamics} simulations, using continuum field variables obtained by \ab*{discrete-to-continuous} method. We found that cross-slip leads to more homogeneous and stable dislocation microstructures within which dislocations are able to remain stable closer to free surfaces. Cross-slip also results in more \emph{similar} dislocation microstructures. These findings indicate that the disadvantage of larger computational complexity when including cross-slip in \ab*{discrete dislocation dynamics}
simulations might be offset by requiring fewer realizations to capture
statistical aspects of dislocation microstructures.

The finding that the dislocation microstructure close to surfaces changes significantly is important for analyzing dislocations experimentally via non-destructive methods.
These observations are influenced by the surface and any conclusions drawn from such observations about the impact of dislocation arrangements must be made carefully.
In order to provide better heuristics on the actual changes near the surface, however, we have to perform further analyzes that consider only surface near regions by taking alignment of the surfaces with respect to the loading direction into account, which is out of the current scope of our work presented here.

Our analyzes for the effects of discretization resolution shows that discretized domains with high resolutions contain too much discrete information which is not in line with the intention of obtaining continuous field data. We observed a lower bound for the resolution by the dislocation density distributions: a voxel length should not be smaller than the average dislocation spacing. In addition, cross-slip has an impact on the formation as the peaks are present in coarser discretizations when cross-slip is not activated. This is simply because of the fact that in the microstructures with cross-slip, the total dislocation density increases more and results in lower average dislocation spacing which allows smaller voxel sizes. The fluctuations are important geometrical artifacts that have to be considered in continuum field data calculations as the they can lead to misjudgments of further analyzes and hidden calculation errors in \ab*{continuum dislocation dynamics} simulations.

From a broader perspective, the outcomes of our study are:
\begin{itemize}
	\item We established a method of descriptive statistics for systems of curved dislocations. By using the \ab*{continuum dislocation dynamics} field variables as descriptors for certain microstructural aspects we indirectly could leverage the fact that CDD is based on a statistical coarse graining of systems of discrete dislocations. In other words, our descriptors are strongly based on physics which make them easily interpretable.  
	\item We demonstrated that using these descriptors it is possible to investigate and to discuss situations that otherwise can only be approached by eyeballing or through very coarse measures. In particular the fact that curved dislocations behave entirely different from straight dislocations (one of the reason why 3D \ab*{discrete dislocation dynamics} is such an important method) could so far not properly be accounted for or leveraged.
	\item All of this is a step towards parameterizing and validating continuum simulations methods, ranging from gradient based models up to multislip \ab*{continuum dislocation dynamics} methods of various complexity. Furthermore, in the context of continuum model development, we have now a methodology that helps us to infer how, e.g., additional terms concerning dislocation multiplication could be included. This, however, is still a significant undertaking, which can not be presented in the present manuscript as well.
	\item Last but not least, a similar analysis can also be done with experimental data (at least to some extent). For example, in ~\autocite{zhang2022data}, we extracted the dislocation geometry from in-situ TEM experiments, converted them using \ab*{discrete-to-continuous} and used in particular the curvature to understand details of the “energy landscape” of high-entrop alloys. Thus, \ab*{discrete-to-continuous} can also be seen as a tool to bring experiments and simulations closer together. Furthermore, \ab*{discrete-to-continuous} method has a potential to be extended for providing information on the mechanical fields of a microstructures in the simulations, however it is out of the scope of the current study.

\end{itemize}


\section*{Acknowledgements}
AD, DS and SS acknowledge funding from the European Research Council Starting Grant, “A Multiscale Dislocation Language for Data-Driven Materials Science”, ERC Grant Agreement No. 759419 MuDiLingo.

\section*{Data and code availability}
The data that support the findings of this study will be openly available following an embargo at the following DOI: \url{10.5281/zenodo.7788934}

\section*{Declarations}
On behalf of all authors, the corresponding author states that there is no conflict of interest.

\appendix

\begin{appendices}
	
	\counterwithin{figure}{section}
	
	\section{Total dislocation density fields}
	\textcolor{white}{white}
	\begin{figure}[htp!]
		\centering
		\includegraphics[width=0.8\linewidth]{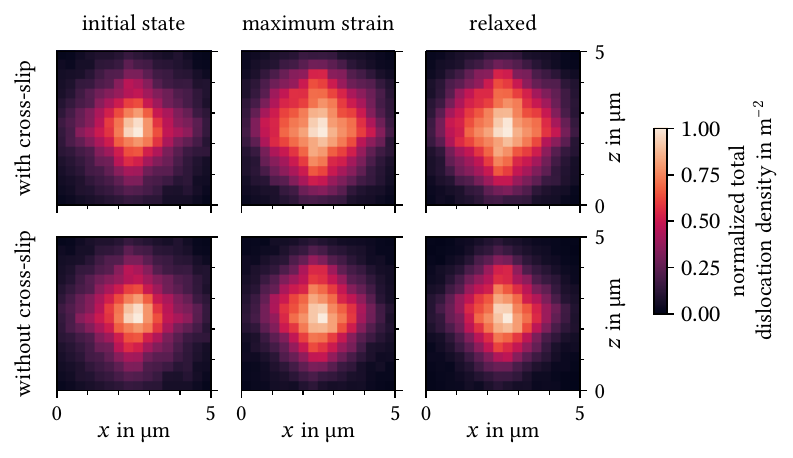}
		\captionbelow{%
			Normalized average total dislocation density per dislocation microstructure set further averaged along the $y$-direction, i.e., the viewing direction of \cref{fig:discrete-dislocation-structures} and this figure.
			The spatial discretization is \num{16} voxels along each direction. Each averaged specimen is normalized by their maximum value of the densities.
		}
		\label{fig:total-dislocation-density-averaged-along-y}
	\end{figure}

	\section{Probability density distributions of total dislocation density}	
		\textcolor{white}{white}    
		\begin{figure}[htp!]
			\centering
			\includegraphics[width=0.8\linewidth]{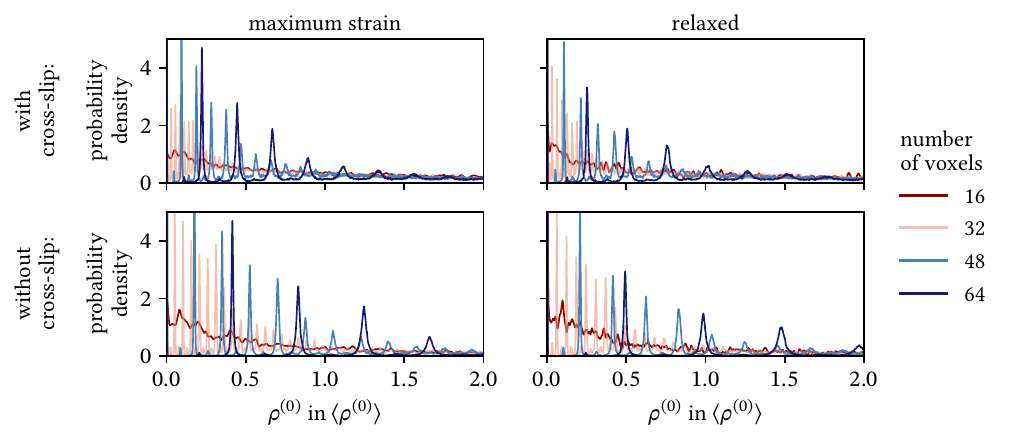}
			\caption{Probability density distributions of the normalized total density for the discretizations by 16, 32, 40 and 64 voxels along each direction. The first and second row show the results from simulations with cross-slip and without cross-slip, respectively. The first and second columns show the results from simulations that are at the maximum strain state and the subsequent relaxed state, respectively.}
			\label{fig:pdfrhomulti_more}
		\end{figure}

		\begin{figure}[htp]
			\centering
			\includegraphics[width=0.8\linewidth]{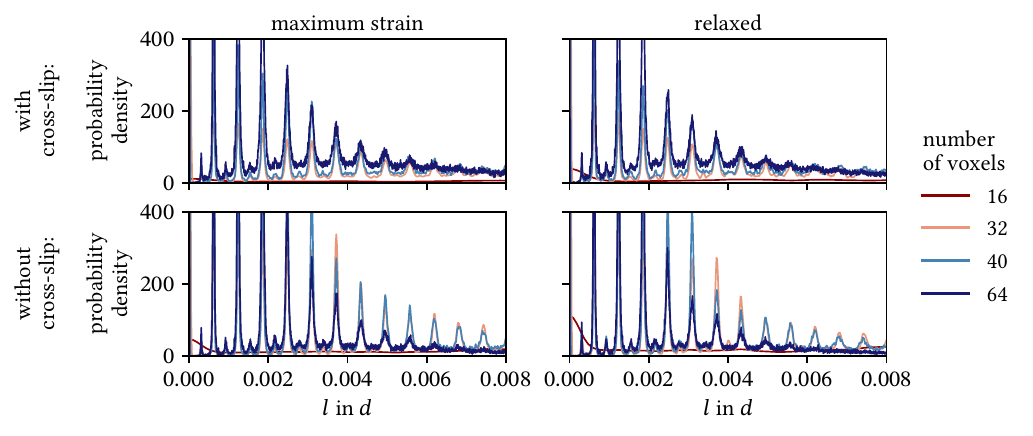}
			\caption{Probability density distributions of the dislocation length in the voxels for the discretizations by 16, 32, 40 and 64 voxels along each direction. $l$ is the total dislocation length in a voxel, and $d=\SI{5.0}{\micro\meter}$ is the specimen edge length, which is used for normalization of the $l$ values. The first and second row show the results from simulations with cross-slip and without cross-slip, respectively. The first and second columns show the results from simulations that are at the maximum strain state and the subsequent relaxed state, respectively.}
			\label{fig:pdflengthmulti_more}
			
		\end{figure}
	
	\section{Curvature density distribution skewness}
		\textcolor{white}{white}
		\begin{figure}[htp]
			\centering
			\includegraphics[width=1\linewidth]{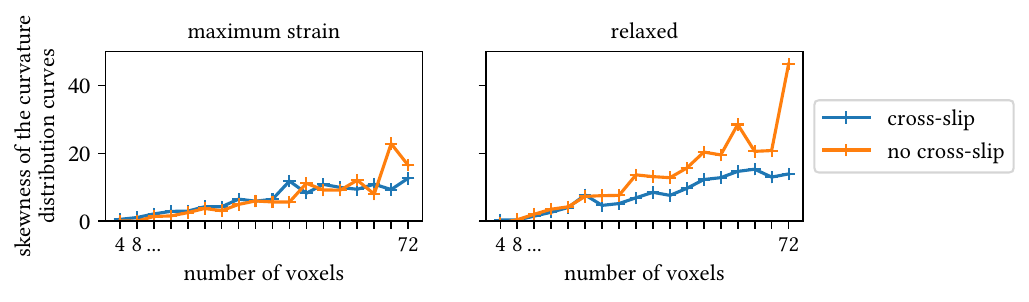}
			\caption{Skewness values calculated from the curves of the curvature distributions for each discretization resolution.}
			\label{fig:skewness}
			
		\end{figure}

\end{appendices}

\newpage
\printbibliography

\end{document}